\documentclass[twocolumn,showpacs,preprintnumbers,amsmath,amssymb]{revtex4}
\usepackage{graphicx}
\usepackage{dcolumn}
\usepackage{putexPRL}

\begin{document}

\preprint{PUPT-2264}

\title{
Colorful horizons with charge in anti-de Sitter space
}

\author{Steven S. Gubser}
\affiliation{Joseph Henry Laboratories, Princeton University, Princeton, NJ 08544}

\date{March 2008}

\begin{abstract}
An abelian gauge symmetry can be spontaneously broken near a black hole horizon in anti-de Sitter space using a condensate of non-abelian gauge fields.  There is a second order phase transition between Reissner-Nordstrom-anti-de Sitter solutions, which are preferred at high temperatures, and symmetry breaking solutions, which are preferred at low temperatures.
\end{abstract}


\maketitle

\section{Introduction}
\label{INTRODUCTION}

I have previously argued that an abelian gauge symmetry can be spontaneously broken by a condensate of a charged scalar that forms near the horizon of a non-extremal charged black hole, presumably indicating some form of superfluidity or superconductivity \cite{Gubser:2005ih,Gubser:2008px}.  Unpublished numerical calculations based on the model proposed in \cite{Gubser:2008px} indicate that ``superconducting'' horizons are thermodynamically preferred below some non-zero critical temperature.  But there are enough parameters in the lagrangians considered in \cite{Gubser:2005ih,Gubser:2008px} that it is challenging to characterize the degree of universality or robustness of such numerical results.  I will therefore consider an analogous phenomenon in a theory whose lagrangian is mostly determined by symmetry principles:
 \eqn{GravityPlusYM}{
  I = {1 \over 2\kappa^2} 
   \int d^4 x \, \sqrt{-g} {\cal L}
 }
where
 \eqn{Ldef}{
  {\cal L} = R + {6 \over L^2} - {1 \over 4} (F_{\mu\nu}^a)^2 \,,
 }
and
 \eqn{FmunuDef}{
  F_{\mu\nu}^a = \partial_\mu A_\nu^a - \partial_\nu A_\mu^a + 
    g \epsilon^{abc} A_\mu^b A_\nu^c
 }
is the field strength of an $SU(2)$ gauge field.  I use mostly plus signature, and $\epsilon^{abc}$ is the totally antisymmetric tensor with $\epsilon^{123}=1$.  I aim to persuade the $A_\mu^1$ and $A_\mu^2$ gauge bosons to leap out of the horizon and condense near it, thereby breaking a $U(1)$ symmetry associated with $A_\mu^3$.  This should be possible if the gauge coupling $g$ is large enough.

It is convenient to represent the gauge field as a matrix-valued one form:
 \eqn{Arep}{
  A = A_\mu^a \tau^a dx^\mu \,,
 }
where $\tau^a = \sigma^a/2i$, so that
 \eqn{tauCommute}{
  [\tau^a,\tau^b] = \epsilon^{abc} \tau^c \,.
 }
(By $\sigma^a$ I mean the usual Pauli matrices.)  I will restrict attention to the following ansatz:
 \begin{subequations}\label{BHansatz}
 \begin{align}
  ds^2 &= e^{2a} \left( -h dt^2 + dx_1^2 + dx_2^2 \right) + 
    {dr^2 \over e^{2a} h}  \label{BHlineElement} \\
  A &= \Phi \tau^3 dt + w (\tau^1 dx_1 + \tau^2 dx_2) \,,
   \label{BHgaugeField}
 \end{align}
 \end{subequations}
where $a$, $h$, $\Phi$, and $w$ are functions only of $r$.  The gauge field \eno{BHgaugeField} is a slight simplification of the ansatz considered in \cite{Bartnik:1988am,Bizon:1990sr}.  A substantial literature has grown up around similar solutions of Einstein-Yang-Mills theory: see \cite{Volkov:1998cc,Winstanley:2008ac} for reviews.

The electrostatic potential $\Phi$ must vanish at the horizon for $A$ to be well-defined as a one-form, but I do not require it to vanish at infinity: thus the black hole can carry charge under the $U(1)$ gauge symmetry generated by $\tau^3$.

The condensate $w(\tau^1 dx_1 + \tau^2 dx_2)$ breaks the $U(1)$ rotation symmetry in the $x_1$-$x_2$ plane as well as the $U(1)$ gauge symmetry generated by $\tau^3$, but it preserves a combination of the two.  I will require $w$ to be normalizable in the sense of making a finite contribution to the norm \footnote{The norm \eno{NormDef} can be made positive definite by passing to a gauge where $A^a_0=0$, but then one has the problem that the norm for a static electric field grows linearly with time.  Such issues do not affect the question of whether $w$ makes a finite contribution to $||A||$.}
 \eqn{NormDef}{
  ||A||^2 \equiv \int_{r_H}^\infty dr \, 
    \sqrt{-g} \, g^{\mu\nu} A^a_\mu A^a_\nu \,,
 }
where $r_H$ is the location of the horizon.  Normalizability of $w$ is what it will mean for the condensate to form ``near'' the horizon.  It is an appropriate requirement in the context of studying spontaneous symmetry breaking.

In summary: the only conserved quantities associated with the black hole should be its mass density and its electric $\tau^3$ charge density; and $w$, if it is non-zero, is a condensate whose presence spontaneously breaks $U(1)$ of spatial rotations times $U(1)$ of $\tau^3$ into a diagonal subgroup.  I want to know when this spontaneous symmetry breaking occurs.

\begin{widetext}
\section{Symmetry breaking solutions}

Plugging \eno{BHlineElement} and \eno{BHgaugeField} into the equations of motion following from \eno{Ldef} results in four second order equations of motion:
 \begin{subequations}\label{eoms}
 \begin{eqnarray}
  \label{eoms1}
  a'' + a'^2 + {1 \over 2} e^{-2a} w'^2 + 
    {g^2 \over 2} e^{-6a} {\Phi^2 w^2 \over h^2} = 0  \\
  \label{eoms2}
  h'' + 4a'h' - e^{-2a} \Phi'^2 - g^2 e^{-6a} {\Phi^2 w^2 \over h} +
    e^{-2a} h w'^2  - e^{-2a} h w'^2 - g^2 e^{-6a} w^4 = 0  \\
  \label{eoms3}
  \Phi'' + 2a'\Phi' - 2g^2 e^{-4a} {\Phi w^2 \over h} = 0  \\
  \label{eoms4}
  w'' + \left( 2a' + {h' \over h} \right) w' + 
   g^2 e^{-4a} \left( {\Phi^2 w \over h^2} - {w^3 \over h} \right)
     = 0
 \end{eqnarray}
 \end{subequations}
 \begin{table}[b]
 \caption{Scaling symmetries of the equations of motion \eno{eoms}, the constraint \eno{constraint}, of series coefficients appearing in \eno{HorizonSeries} and \eno{FarField}, and of the thermodynamic variables appearing in \eno{ThermoQuantities}.\label{ChargeAssignments}}
 \begin{ruledtabular}
 \begin{tabular}{c|>{$}c<{$}|>{$}c<{$}|>{$}c<{$}|>{$}c<{$}|>{$}c<{$}|>{$}c<{$}|>{$}c<{$}|>{$}c<{$}|>{$}c<{$}|>{$}c<{$}|>{$}c<{$}|>{$}c<{$}|>{$}c<{$}|>{$}c<{$}|>{$}c<{$}|>{$}c<{$}|>{$}c<{$}|>{$}c<{$}|>{$}c<{$}|>{$}c<{$}|>{$}c<{$}|>{$}c<{$}|>{$}c<{$}|>{$}c<{$}|>{$}c<{$}|>{$}c<{$}}
  & dt & d\vec{x} & dr & e^a & e^{a_0} & h & h_1 & H_0 & H_3 & \Phi & \Phi_1 & p_0 & p_1 & w & w_0 & W_1 & L & g & \kappa & I & \epsilon & s & T & \rho & \mu & J \\ \hline 
 A & 0 & 1 & 0 & -1 & -1 & 2 & 2 & 2 & -1 & 0 & 0 & 0 & -1 & -1 & -1 & -2 & 0 & 0 & 0 & 0 & -3 & -2 & -1 & -2 & -1 & -2  \\
 B & 1 & 1 & -1 & -1 & -1 & 0 & 1 & 0 & -3 & -1 & 0 & -1 & -2 & -1 & -1 & -2 & 0 & 0 & 0 & 0 & -3 & -2 & -1 & -2 & -1 & -2  \\
 C & 1 & 1 & 1 & 0 & 0 & 0 & -1 & 0 & 0 & 0 & -1 & 0 & 0 & 0 & 0 & 0 & 1 & -1 & 1 & 0 & -3 & -2 & -1 & -2 & -1 & -2  \\
 D & 0 & 0 & 0 & 0 & 0 & 0 & 0 & 0 & 0 & 0 & 0 & 0 & 0 & 0 & 0 & 0 & 0 & 0 & 1 & -2 & -2 & -2 & 0 & -2 & 0 & -2
\end{tabular}
 \end{ruledtabular}
 \vskip0.1in
 \end{table}
together with a zero-energy constraint,
 \eqn{constraint}{
  12 e^{2a} a'^2 + 4 e^{2a} {h' \over h} a' - 
   {12 \over L^2 h} + {\Phi'^2 \over h} - 
   2w'^2 - 2g^2 e^{-4a} {\Phi^2 w^2 \over h^2} + 
   g^2 e^{-4a} {w^4 \over h} = 0 \,.
 }
\end{widetext}
The metric and the gauge field must be regular at the horizon, which I assume to be at $r=0$.  This implies series expansions around $r=0$ of the form
 \eqn{HorizonSeries}{
  a &= a_0 + a_1 r + a_2 r^2 + \ldots  \cr
  h &= h_1 r + h_2 r^2 + \ldots  \cr
  \Phi &= \Phi_1 r + \Phi_2 r^2 + \ldots  \cr
  w &= w_0 + w_1 r + w_2 r^2 + \ldots \,.
 }
It is also straightforward to obtain the asymptotic behavior at large $r$:
 \eqn{FarField}{
  a &= \log {r \over L} + \alpha_0 + \ldots  \cr
  h &= H_0 + H_3 e^{-3a} + H_4 e^{-4a} + \ldots \cr
  \Phi &= p_0 + p_1 e^{-a} + \ldots \cr
  w &= W_1 e^{-a} + \ldots \,.
 }
Once the six quantities $a_0$, $h_1$, $\Phi_1$, $w_0$, $g$, and $L$ are specified, all the other coefficients in \eno{HorizonSeries} may be computed, and one can then use the series expansions to specify Cauchy data for the equations of motion \eno{eoms} at some radius $r$ slightly outside the horizon.  Of these six parameters, only three are meaningful, because the equations \eno{eoms} and \eno{constraint} have three scaling symmetries that act non-trivially on the parameters.  These symmetries are summarized in the first three rows of table~\ref{ChargeAssignments}, where assigning a charge $\alpha$ to a quantity $X$ means
 \eqn{Xtransform}{
  X \to \lambda^\alpha X \,.
 }
Recall that $w$ is required to be normalizable at infinity.  This requirement is imposed in the last equation of \eno{FarField}, but it is not the typical behavior of $w$: usually there is a constant term at large $r$, or else there is a singularity at finite $r$.  The solutions with normalizable $w$ form a co-dimension one locus in the three-dimensional space of parameters modulo scaling symmetries.  This locus includes as one branch the well-known Reissner-Nordstrom-anti-de Sitter solutions (hereafter RNAdS), which have the form \eno{BHansatz} with $w=0$: see for example \cite{Hartnoll:2007ai}.

The energy density, entropy density, temperature, chemical potential, and charge density, as well as an order parameter, $J$, to be discussed further below, can be read off as follows:
 \eqn{ThermoQuantities}{\seqalign{\span\TL & \span\TR & \qquad\qquad \span\TL & \span\TR}{
  \epsilon &= -{H_3 \over \kappa^2 L H_0} &
  s &= {2\pi \over \kappa^2} e^{2a_0}  \cr
  \mu &= {p_0 \over 2L\sqrt{H_0}} & 
  T &= {1 \over 4\pi} e^{2a_0} {h_1 \over \sqrt{H_0}}  \cr
  \rho &= -{p_1 \over \kappa^2 \sqrt{H_0}} &
  J &= {W_1 L \over \kappa^2} \,.
 }}
Energy density, temperature, and chemical potential are measured in reference to the Killing time $\sqrt{H_0} t$ rather than $t$ itself.  
The quantities in \eno{ThermoQuantities} can be regarded as characterizing a thermal state of a dual conformal field theory, along the lines of the gauge-string duality \cite{Maldacena:1997re,Gubser:1998bc,Witten:1998qj}.  Their normalizations accord with the conventions of \cite{Hartnoll:2007ai}.  
All dependence on $\kappa$ can be removed by defining
 \eqn{HattedDefs}{\seqalign{\span\TL & \span\TR & \qquad\qquad \span\TL & \span\TR}{
  \hat\epsilon &= {\kappa^2 \epsilon \over (2\pi)^3 L^2} &
  \hat{s} &= {\kappa^2 s \over (2\pi)^3 L^2}  \cr
  \hat\rho &= {\kappa^2 \rho \over (2\pi)^3 L^2} &
  \hat{J} &= {\kappa^2 J \over (2\pi)^3 L^2} \,,
 }}
where the factors of $2\pi$ are included for later convenience.
Any relation among the quantities in \eno{ThermoQuantities} and \eno{HattedDefs} must be expressible in terms of ratios which are invariant under the scaling symmetries summarized in table~\ref{ChargeAssignments}.  For example, the RNAdS solutions have
 \eqn{EOS}{
  {\hat\epsilon \over \hat{s}^{3/2}} = 
   1 + {\pi^2 \hat\rho^2 \over \hat{s}^2} \,.
 }

When $w\neq 0$, one may ask what fraction $q$ of the electric charge density is carried by the non-abelian gauge bosons outside the horizon.  The ratio of the flux of the $\tau^3$ electric field through the horizon to the flux at infinity is $1-q$.  Therefore,
 \eqn{qFormula}{
  q = 1 + L e^{2a_0} \sqrt{H_0} {\Phi_1 \over p_1} \,.
 }
$q$ by itself is invariant under the scaling symmetries.

The conformal field theory dual to the $AdS_4$ background under consideration has currents $J^a_m$ satisfying an $SU(2)$ current algebra, where $m$ runs over the $t$ and $x_i$ directions.  The symmetry breaking that arises from non-zero $w$ corresponds to expectation values
 \eqn{expectJ}{
  \langle J^a_i \rangle \propto J \delta^a_i \,,
 }
where $i$ runs over the values $1,2$.  The tensor $\delta^a_i$ exhibits the locking of a spatial $U(1)$ and a gauge $U(1)$.  \eno{expectJ} describes a form of long-range order which infrared fluctuations probably destroy; however, fluctuations are suppressed in the supergravity approximation, where $\kappa \ll L$ \footnote{Infrared fluctuations can also be regulated using finite volume constructions, such as toroidal compactification in the $x_1$-$x_2$ directions, or a generalization to spherically symmetric black holes in global $AdS_4$.  I thank D.~Huse for discussions on these points.}.  I will therefore ignore them.

\section{Summary of results}
\label{Summary}

 \begin{figure*}
  \begin{center}
   \begin{picture}(500,400)(0,0)
    \put(0,230){\includegraphics[width=2.8in]{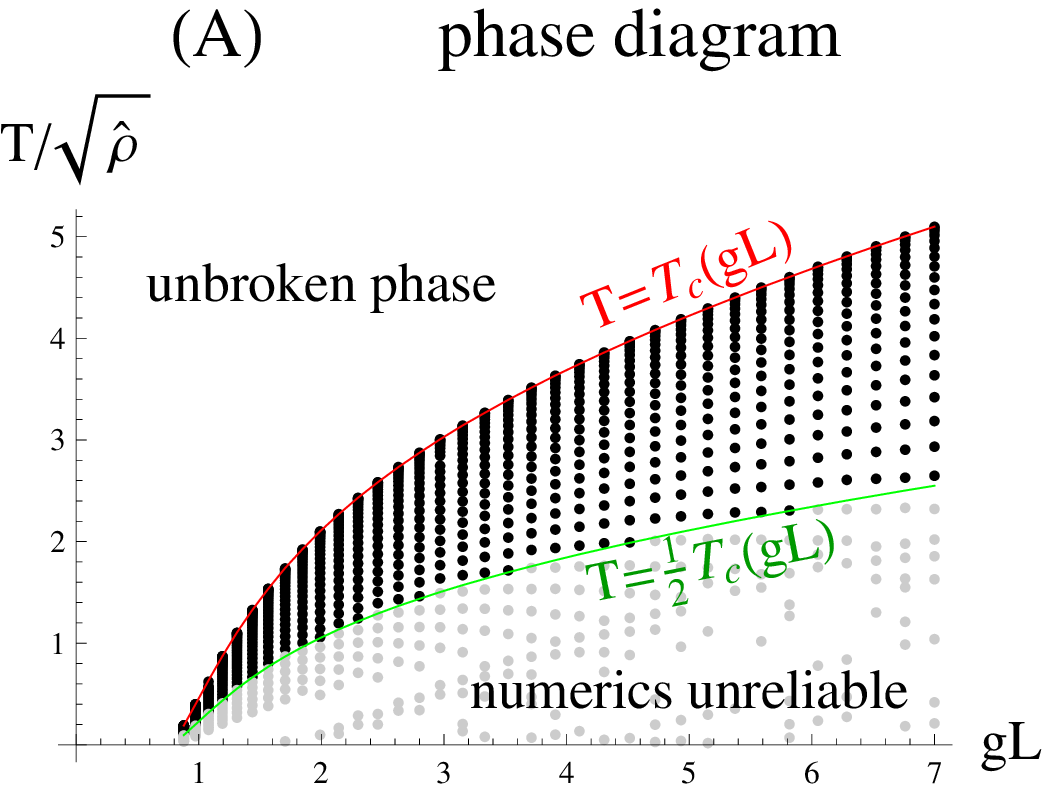}}
    \put(250,210){\includegraphics[width=3in]{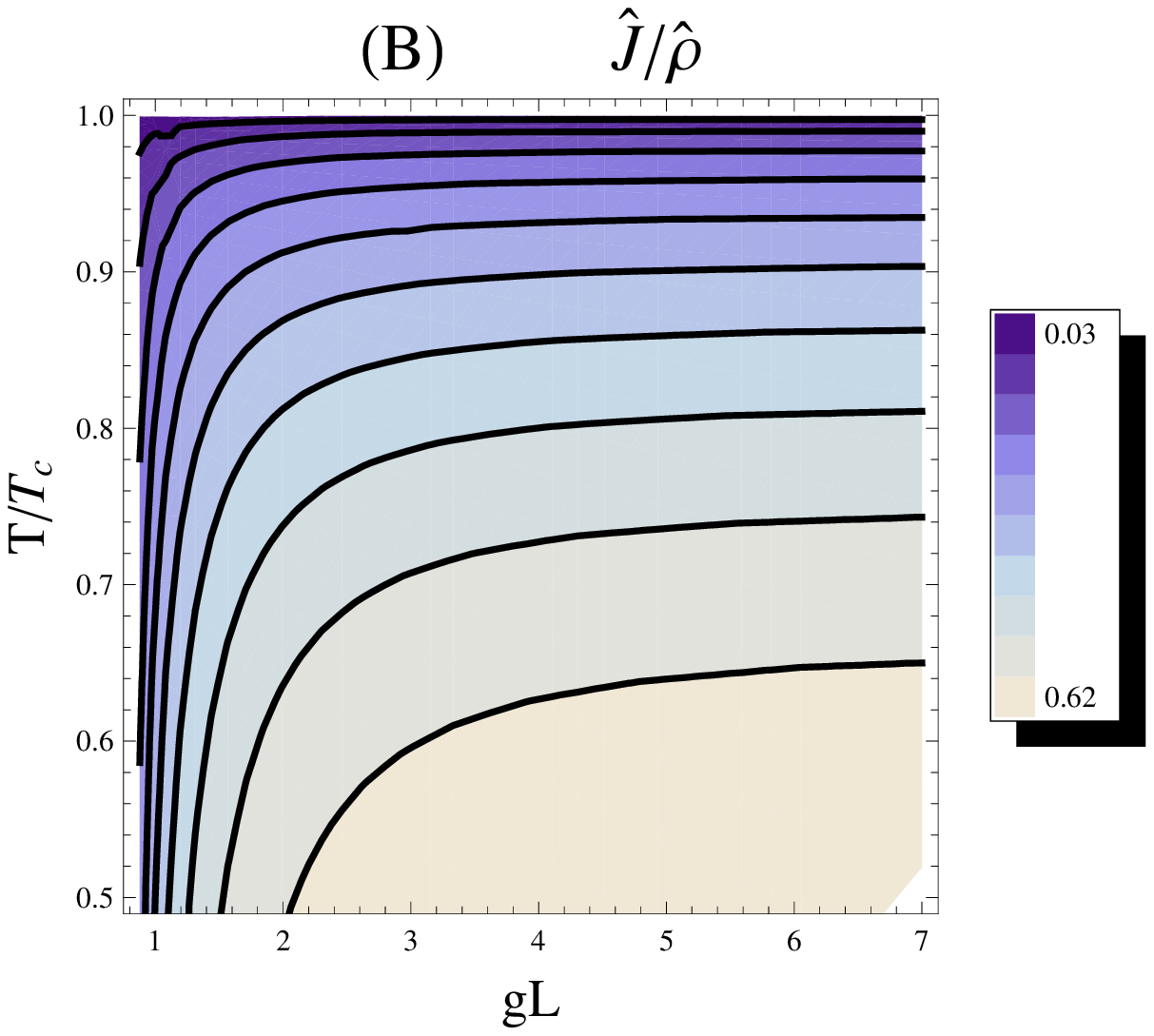}}
    \put(0,0){\includegraphics[width=3in]{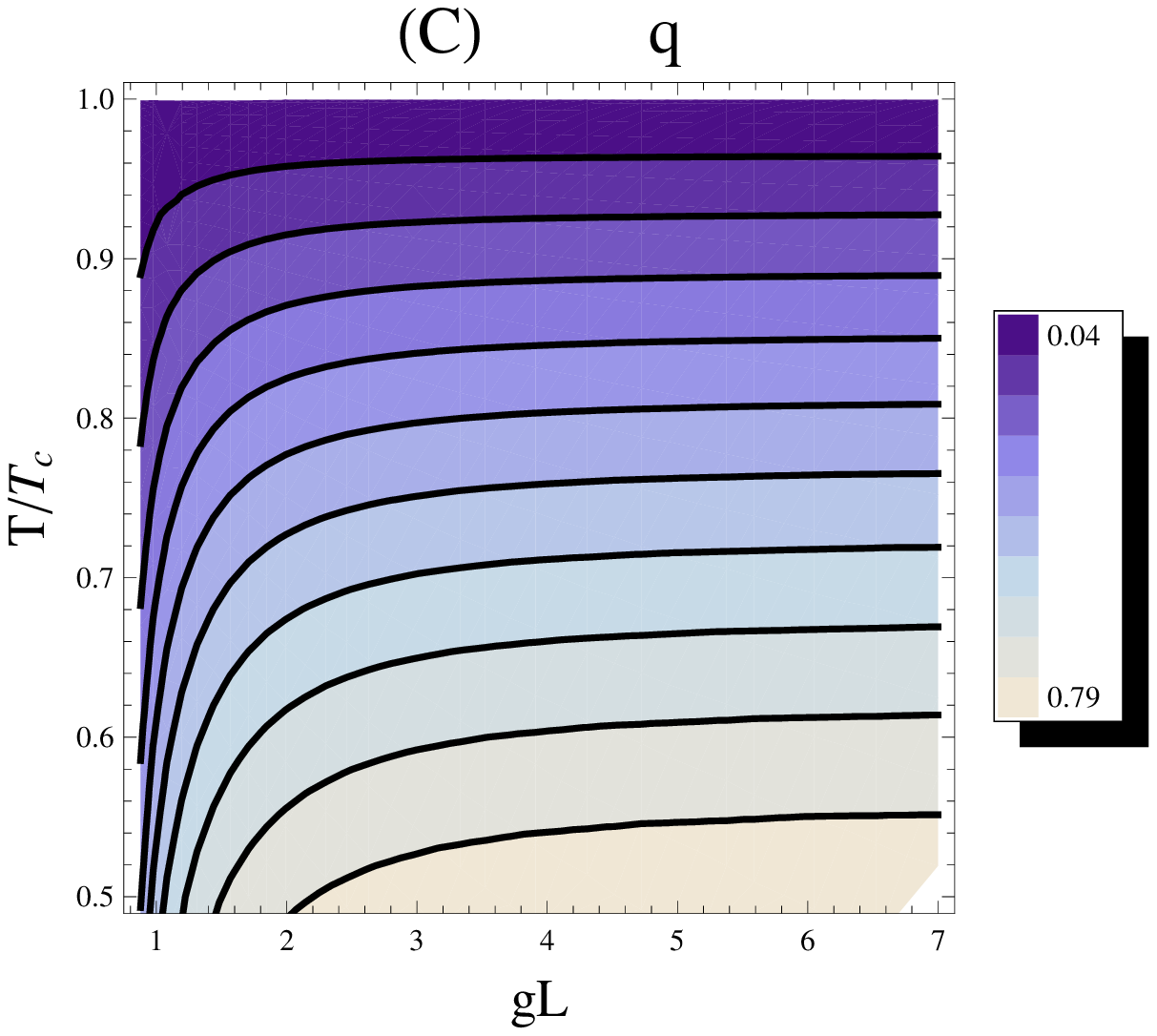}}
    \put(250,0){\includegraphics[width=3in]{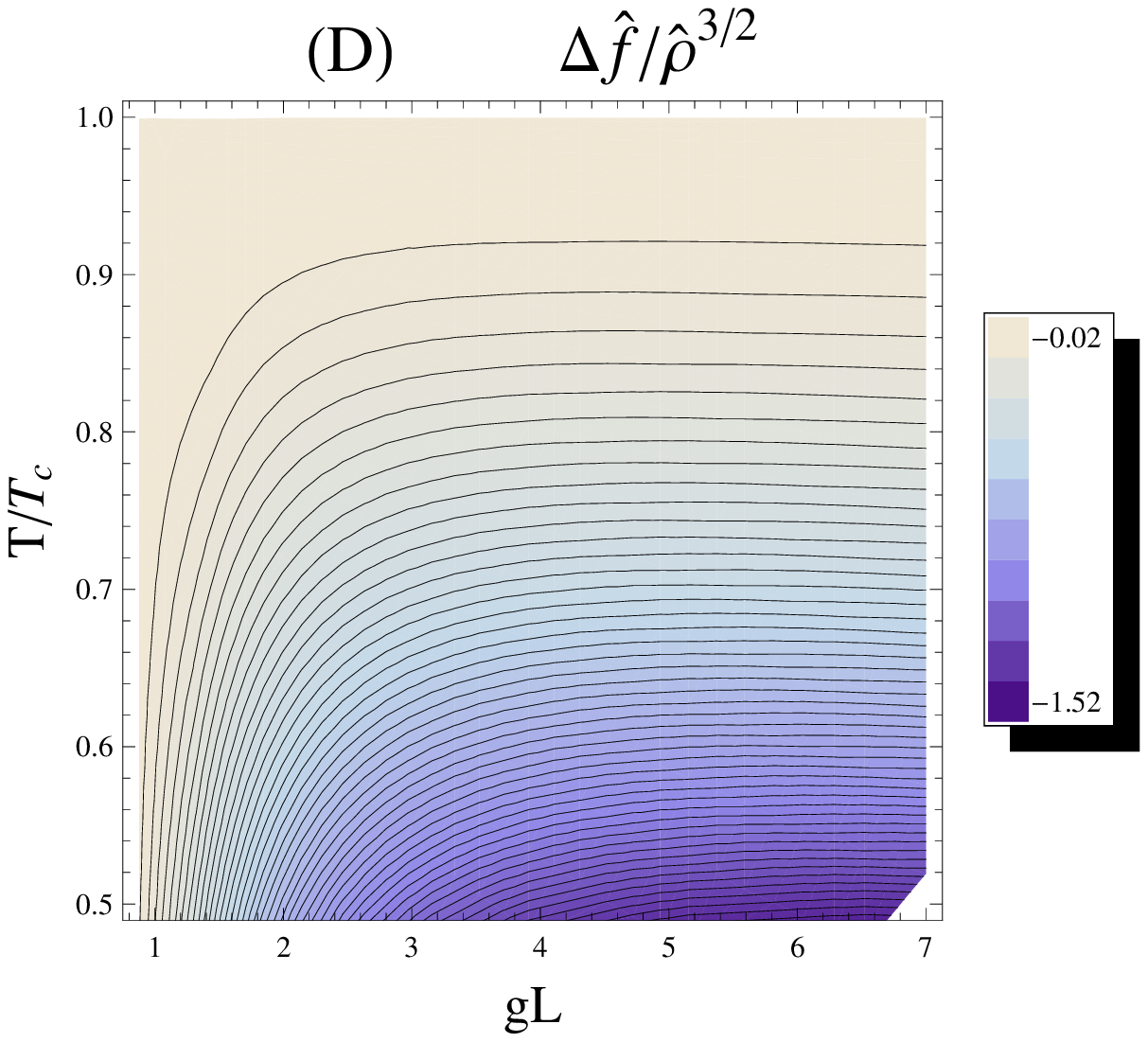}}
   \end{picture}
  \end{center}
 \caption{\label{SummaryPlots}
(A) The phase diagram.  Each point corresponds to a numerically constructed black hole solution.  (B) $\hat{J}/\hat\rho$ as a function of $T/T_c$.  (C) $q$ as a function of $T/T_c$.  (D) $\Delta\hat{f}/\hat\rho^{3/2}$ as a function of $T/T_c$.
}
\end{figure*}

As explained following \eno{Xtransform}, there is a two-parameter family of black hole solutions with a non-vanishing, normalizable condensate, once all possible scaling symmetries are used.  A convenient choice of parameters is $gL$ and $T/\sqrt{\hat\rho}$.  Other scaling invariants include $\hat{J}/\hat\rho$, $q$, and $\Delta \hat{f}/\hat\rho^{3/2}$, where
 \eqn{DeltaFhatDef}{
  \Delta\hat{f} = {\kappa^2 \Delta f \over (2\pi)^3 L^2} \,.
 }
Here $\Delta f$ is the difference in the free energy density, $\epsilon-Ts$, between a symmetry breaking solution and the RNAdS solution with the same $T$ and $\rho$.  $\Delta f < 0$ means that the symmetry breaking solution is preferred.

I studied solutions to \eno{eoms} numerically for about 1600 different choices of parameters: see figure~\ref{SummaryPlots}.  I restricted attention to solutions with $w$ everywhere positive.  Solutions exist in which $w$ has nodes, but they are probably always thermodynamically disfavored because spatial oscillations in $w$ increase energy density.  A shooting strategy was employed to implement the constraint of normalizability on $w$.  Solutions with non-zero, normalizable $w$ were found only for $T$ less than a critical temperature $T_c$.  This $T_c$ is, within numerical error, the temperature at which the RNAdS solution admits a static linearized perturbation, with $w$ non-zero but infinitesimally small.  The shooting algorithm was designed to work well near $T_c$.  In practice, it worked well for $T>T_c/2$.

$T_c$ goes to zero at $g = g_c \approx 0.8/L$ and does not appear to have singular behavior there; however, the numerical problem becomes more difficult at small $T_c$, so the value of $g_c$ should be regarded as approximate.  For $g \gsim 4g_c$, scaled thermodynamic quantities such as $\hat{J}/\hat\rho$, $q$, and $\Delta\hat{f}/\hat\rho^{3/2}$ exhibit nearly universal behavior as functions of $T/T_c$ from $T=T_c$ down at least to ${1 \over 2} T_c$.  This universality is related to a large $g$ limit where the back-reaction of the gauge field on the metric can be neglected.

For $0.75 T_c < T < T_c$, I found good fits to the following scaling forms:
 \eqn{ScalingForms}{
  q = q_1 t \qquad {\hat{J} \over \hat\rho} = j_{1/2} \sqrt{t} \,,
 }
where $t = 1-T/T_c$, and $q_1$ and $j_{1/2}$ are positive and depend on $gL$.  More approximate fits can be found over the same temperature range to the scaling form
 \eqn{FScaling}{
  {\Delta\hat{f} \over \hat\rho^{3/2}} = -f_2 t^2 \,,
 }
where $f_2$ is positive and depends on $gL$.  A few solutions close to $T_c$ appeared to have $\Delta f$ very slightly positive.  But closer examination of some of these solutions using highly accurate numerics showed that in fact they have $\Delta f < 0$.

In conclusion: there is a second order phase transition, with simple critical exponents, from Reissner-Nordstrom-anti-de Sitter solutions to solutions with normalizable non-abelian condensates.  The critical temperature scales as the square root of charge density and also depends on the gauge coupling.  The solutions with condensates break an abelian gauge symmetry, suggesting some link or analogy to superconductivity.  An analogy has been already suggested between charged anti-de Sitter black holes without symmetry breaking condensates and the pseudogap state of high $T_c$ materials \cite{Herzog:2007ij,Hartnoll:2007ih,Hartnoll:2007ip,Hartnoll:2008hs}.  It would be desirable to test the validity of this analogy more closely, and to extend it if possible.  Given the way spatial rotations enter into the description of the condensate \eno{expectJ}, it might be possible in some variant of the construction I have exhibited to find an analog of $p$-wave or $d$-wave superconductivity.  Any such analog with high $T_c$ superconductivity should be viewed with caution if not skepticism, because the underlying degrees of freedom of duals to $AdS_4$ vacua are typically large $N$ gauge theories, seemingly distant from semi-realistic constructions like the Hubbard model.  And yet, one might hope that dynamics related to the global symmetries (usually $U(2)$ or $SO(4)$, and sometimes larger groups) of the Hubbard model and its relatives can be mimicked, to some extent, by non-abelian gauge fields on the gravity side of the gauge-string duality.

\begin{acknowledgments}

I thank C.~Herzog for discussions.  This work was supported in part by the Department of Energy under Grant No.\ DE-FG02-91ER40671 and by the NSF under award number PHY-0652782.

\end{acknowledgments}

\appendix

\section{Static perturbations}

The purpose of the appendices is to consider two limiting cases of the equations \eno{eoms}: the first, in this appendix, is the case where $w$ is treated as a small perturbation to an RNAdS solution; and the second, in appendix~\ref{STRONG}, is the case where the entire non-abelian gauge field is treated as a small perturbation to the $AdS_4$-Schwarzschild geometry.

\begin{figure*}
  \centerline{\includegraphics[width=7in]{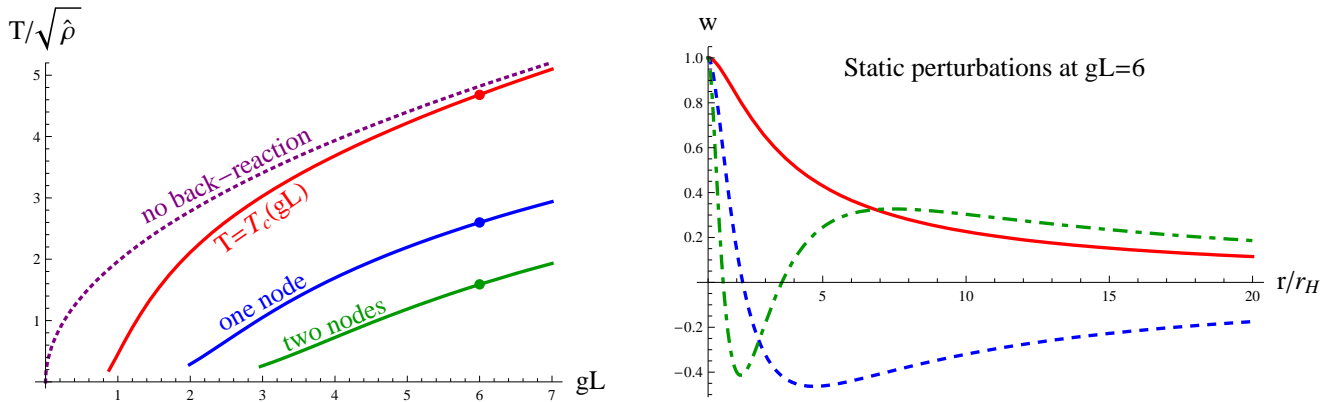}}
  \caption{\label{StaticSolns}
Left: Temperatures where static linear perturbations of the RNAdS solution arise.  The curve labeled ``no back-reaction'' is given analytically by \eno{FoundTc}.  It comes from neglecting the back-reaction of $\Phi$ as well as of $w$ on the metric.  As discussed in section~\ref{STRONG}, neglecting back-reaction is a good approximation when $gL$ is large.  Right: Static normalizable perturbations, scaled so that $w=1$ at the horizon.  The solution that is everywhere positive corresponds to the dot at $gL=6$ on the $T=T_c$ curve in the left-hand plot, and the other solutions correspond to the dots on the other curves.
}
 \end{figure*}

The RNAdS background is
 \eqn{RNAdS}{
  ds^2 &= -fdt^2 + {r^2 \over L^2} d\vec{x}^2 + {dr^2 \over f}  \cr
  f &= -{\epsilon L^2 \kappa^2 \over r} + 
    {\rho^2 \kappa^4 L^2 \over 4r^2} + {r^2 \over L^2}  \cr
  \Phi &= \rho L \kappa^2 \left( {1 \over r_H} - {1 \over r} 
    \right) \,,
 }
with $w=0$.  By definition, $r_H$ is the most positive zero of $f$.  If $w$ is treated as a static linear perturbation, then the only equation of motion we must solve is \eno{eoms4} with the $w^3$ term neglected:
 \eqn{LinearizedW}{
  w'' + {f' \over f} w' + {g^2\Phi^2 \over f^2} w = 0 \,.
 }
Although \eno{LinearizedW} does not appear to be solvable in terms of known functions, it is easy to make a series expansion near the horizon:
 \eqn{wHorizon}{
  w = w_0 + w_1 (r-r_H) + w_2 (r-r_H)^2 + \ldots \,.
 }
Once $w_0$ is specified, all other coefficients in \eno{wHorizon} can be computed.  Because \eno{wHorizon} is linear, it suffices to consider the case $w_0=1$, even though our real interest is in solutions with very small $w$.  In addition to \eno{wHorizon}, there is also a solution that has a logarithmic singularity at the horizon, but I discard it based on the usual argument that the horizon is a non-singular location.  Far from the horizon, the typical behavior of $w$ is for it to be asymptotically constant, but by adjusting $r_H/L$ one can arrange for $w$ to have $1/r$ falloff, which is normalizable in the sense described around \eno{NormDef}.  The discretely many allowed values of $r_H/L$ depend on $\hat\epsilon$, $\hat\rho$, $g$, and $L$, but because of scaling symmetries (essentially, the ones in table~\ref{ChargeAssignments}), dimensionless quantities such as $T/\sqrt{\hat\rho}$ depend only on $gL$.  Thus one may convert the allowed values of $r_H/L$ into allowed values of $T/\sqrt{\hat\rho}$ at some specified $gL$.  The largest allowed value of $T/\sqrt{\hat\rho}$ corresponds to a solution where $w$ has no zeroes, and in this case the temperature is precisely the $T_c$ discussed in the paragraph following \eno{DeltaFhatDef}.  The next-to-largest allowed value of $T/\sqrt{\hat\rho}$ corresponds to a solution with one node; the next-to-next-to-largest to two nodes; and so forth.  See figure~\ref{StaticSolns}.

The last term in \eno{wHorizon} shows that the condensate develops for essentially the same reason as in the scalar case \cite{Gubser:2008px}: $w$ acquires a negative effective mass squared near the horizon.  When the $w^3$ term in \eno{eoms4} becomes significant, it decreases or limits the tendency of $w$ to condense.  An analogous situation in the lagrangian studied in \cite{Gubser:2008px} would be to have a positive $|\psi|^4$ term in the scalar potential, similarly limiting the tendency of the charged scalar $\psi$ to condense.

\section{Strong coupling limit}
\label{STRONG}

 \begin{figure*}
   \vskip-4in
   \centerline{\includegraphics[width=7.5in]{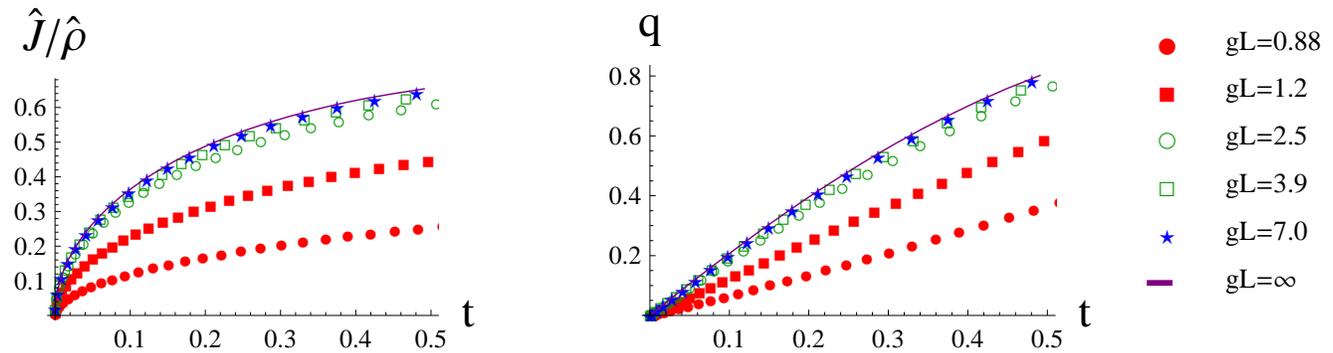}}
   \vskip-0.2in
 \caption{\label{qJCompare}
Plots of $\hat{J}/\hat\rho$ and $q$ as a function of $t = 1-T/T_c$ for several values of $gL$.  The $gL=\infty$ curve is obtained from solutions to \eno{YMscaled}, where back-reaction of the gauge fields on the metric is neglected.
}
\end{figure*}

The large $gL$ limit is simple because the gauge field doesn't back-react significantly on the metric \footnote{I am informed by C.~Herzog that the model proposed in \cite{Gubser:2008px} admits a similar limit \cite{HHHtoappear}.}.  To see this, note that upon scaling $A \to A/g$ and $F \to F/g$, the action for the gauge fields is just ${1 \over 4g^2} (F_{\mu\nu}^a)^2$.  The stress tensor acquires a similar factor of $1/g^2$, and it is this factor that suppresses the back-reaction.  Approximate solutions to the full equations of motion \eno{eoms} and the constraint \eno{constraint} can therefore be constructed by starting with the $AdS_4$-Schwarzschild line element,
 \eqn{AdSSch}{
  ds^2 = {r^2 \over L^2} \left[ -(1-r_H^3/r^3) dt^2 + 
    d\vec{x}^2 \right] + {L^2 \over r^2} {dr^2 \over 1-r_H^3/r^3} \,,
 }
and then solving the Yang-Mills equations in that background.  Let us set $r_H=1$.  Then \eno{AdSSch} has the form \eno{BHlineElement} with
 \eqn{ahChoice}{
  a = \log {r \over L} \qquad h = 1 - {1 \over r^3} \,.
 }
Let us also define
 \eqn{wPhiScaled}{
  \tilde\Phi = gL^2 \Phi \qquad
  \tilde{w} = gL^2 w \,.
 }
Then the Yang-Mills equations \eno{eoms3} and \eno{eoms4} become
 \eqn{YMscaled}{
  \tilde\Phi'' + {2 \over r} \tilde\Phi' - 
    {2 \over r (r^3-1)} \tilde{w}^2 \tilde\Phi &= 0  \cr
  \tilde{w}'' + {1+2r^3 \over r (r^3-1)} \tilde{w} + 
    {r^2 \over (r^3-1)^2} \tilde\Phi^2 \tilde{w} - 
    {2 \over r (r^3-1)} \tilde{w}^3 &= 0 \,.
 }
A charming feature of \eno{YMscaled} is that there are no free parameters in the equations.  Factors of $g$ are absent because the definitions \eno{wPhiScaled} include the aforementioned scaling $A \to A/g$.  It is perhaps surprising that one also needs explicit factors of $L$ in \eno{wPhiScaled} to avoid them in \eno{YMscaled}.  This is because of the choice of radial variable in \eno{AdSSch}: if I had used $z = L^2/r$, such factors would not be needed, because then $L$ enters into the line element only as an overall factor, and overall factors don't effect conformally invariant equations like the classical Yang-Mills equations of motion.

The equations \eno{YMscaled} cannot be solved analytically, but they are straightforward to solve numerically, starting with expansions
 \eqn{MiniHorizonSeries}{
  \tilde\Phi &= \tilde\Phi_1 (r-1) + \tilde\Phi_2 (r-1)^2 + \ldots
    \cr
  \tilde{w} &= \tilde{w}_0 + \tilde{w}_1 (r-1) + \ldots
 }
near the horizon and 
 \eqn{MiniFarField}{
  \tilde\Phi &= \tilde{p}_0 + {\tilde{p}_1 \over r} + \ldots  \cr
  \tilde{w} &= {\tilde{W}_1 \over r} + \ldots
 } 
far from it.  The only free parameters are $\tilde\Phi_1$ and $\tilde{w}_0$: all higher coefficients of \eno{MiniHorizonSeries} can be determined in terms of them.  The normalizable form of $\tilde{w}$ shown in \eno{MiniFarField} is not the typical behavior: instead, $\tilde{w}$ usually asymptotes to a non-zero constant at large $r$.  Requiring normalizable $w$ amounts to a condition on $\tilde\Phi_1$ and $\tilde{w}_0$, and one winds up with a one-parameter family of allowed solutions to \eno{YMscaled}.  Again I restrict attention to solutions with $\tilde{w}>0$ everywhere on the expectation that solutions where $\tilde{w}$ has nodes are never thermodynamically preferred.

Starting from \eno{wPhiScaled} and \eno{HattedDefs}, it is easy to show that
 \eqn{MiniThermo}{
  {T \over \sqrt{\hat\rho}} = 3 \sqrt{-
    {\pi gL \over 2\tilde{p}_1}} \qquad
  q = 1 + {\tilde\Phi_1 \over \tilde{p}_1} \qquad
  {\hat{J} \over \hat\rho} = -{\tilde{W}_1 \over \tilde{p}_1} \,.
 }
I found that $\tilde{p}_1$ is never greater than about $3.65$, and as this value is approached, $\tilde{w}_0 \to 0$.  Using \eno{MiniThermo}, this translates into a critical temperature
 \eqn{FoundTc}{
  T_c \approx 1.97 \sqrt{gL\hat\rho} \,.
 }
(See figure~\ref{StaticSolns}.)  I constructed solutions to \eno{YMscaled} corresponding to a range of temperatures ${1 \over 2} T_c < T < T_c$.  A summary of their thermodynamic behaviors is shown in figure~\ref{qJCompare}.

At lower temperatures, the numerical problem becomes more difficult, partly because a larger range of $r$ must be integrated over.  Also, for smaller values of $T$, extra solutions exist where $\tilde{w}$ has nodes.  This is problematic because shooting algorithms generally rely on numerically finding zeroes of a merit function, and when there are multiple zeroes, one cannot easily control which of them one will find.

\bibliographystyle{apsrev}
\bibliography{ym}

\end{document}